# Achievement of Minimized Combinatorial Test Suite for Configuration-Aware Software Functional Testing Using the Cuckoo Search Algorithm

**\*Bestoun S. Ahmed: Corresponding Author**
*Software Engineering Department, Engineering College, Salahaddin University-Hawler (SUH), 44002, Erbil – Kurdistan Region, Iraq*

*e-mail: bestoon82@gmail.com*

**Taib Sh. Abdulsamad**
*Statistic & Computer Department, College Of Commerce, University of sulaimani*
*Zanko Street, Sulaimania, Kurdistan Region*
*e-mail:Taib.shamsadin@yahoo.com*

**Moayad Y. Potrus**
*Software Engineering Department, Engineering College, Salahaddin University-Hawler (SUH), 44002, Erbil – Kurdistan Region, Iraq*
*e-mail:* moayad_75@yahoo.com

## Abstract

**Context:** Software has become an innovative solution nowadays for many applications and methods in science and engineering. Ensuring the quality and correctness of software is challenging because each program has different configurations and input domains. To ensure the quality of software, all possible configurations and input combinations need to be evaluated against their expected outputs. However, this exhaustive test is impractical because of time and resource constraints due to the large domain of input and configurations. Thus, different sampling techniques have been used to sample these input domains and configurations.

**Objective:** Combinatorial testing can be used to effectively detect faults in software-under-test. This technique uses combinatorial optimization concepts to systematically minimize the number of test cases by considering the combinations of inputs. This paper proposes a new strategy to generate combinatorial test suite by using cuckoo search concepts.

**Method:** Cuckoo Search is used in the design and implementation of a strategy to construct optimized combinatorial sets. The strategy consists of different algorithms for construction. These algorithms are combined to serve the Cuckoo Search.

**Results:** The efficiency and performance of the new technique were proven through different experiment sets. The effectiveness of the strategy is assessed by applying the generated test suites on a real-world case study for the purpose of functional testing.

**Conclusion:** Results show that the generated test suites can detect faults effectively. In addition, the strategy also opens a new direction for the application of Cuckoo Search in the context of software engineering.

***Keywords*:** Combinatorial testing; Search-based software testing; Cuckoo Search; Test case design techniques; covering array; Test Generation Tools, mutation testing.

## 1. Introduction

Testing is the process of evaluating the functionality of a system to identify any gaps, errors, missing requirements, and other features. This process ensures the sound operation of software[1]. In general, testing is mainly classified as either functional and structural [2, 3]. The former method is referred to as "black box testing," and the latter is called "white box testing" [2-4].

In functional testing, the tester ignores the internal structure of the system-under-test and focuses only on the inputs and expected outputs. The technique serves the overall functionality validation of the system, thereby identifying both valid and invalid inputs from the customer's point of view. Structural testing is used to detect logical errors in software [3]. The tester needs to gather information on the internal structure of the system-under-test and to use information with regard to the data structures and algorithms surrounded by the code [5].

Unlike in structural testing, creating a data set (i.e., test data generation) is an important task in functional testing because of the lack of information about the internal design. Previous studies have reported many test data generation methods. In general, these methods use the available information in software requirement specifications, which provide knowledge about input requirements. The tester considers all possible input domains when selecting test cases for the software-under-test. However, considering all inputs is impossible in many practical applications because of time and resource constraints. Hence, the role of test design techniques is highly important.

A test design technique is used to systematically select test cases through a specific sampling mechanism. This procedure optimizes the number of test cases to obtain an optimum test suite, thereby eliminating the time and cost of the testing phase in software development. Different studies proposed various functional test design techniques, such as equivalence class partitioning, boundary value analysis, and cause and effect analysis via decision tables [3, 6]. In general, the tester aims to use more than one testing method because different faults may be detected when different testing methods are used. However, with the vast growth and development of software systems and their configurations, the probability of the occurrence of faults has increased because of the combinations of these configurations, particularly for highly configurable software systems. Traditional test design techniques are useful for fault discovery and prevention. However, such techniques cannot detect faults that are caused by the combinations of input components and configurations [7]. Considering all configuration combinations leads to exhaustive testing, which is impossible because of time and resource constraints [2, 8, 9].

Strategies have been developed in the last 20 years to solve the above problem. Among these strategies, combinatorial testing strategies are the most effective in designing test cases for this problem. These strategies facilitate search and generate a set of tests, thereby forming a complete test suite that covers the required combinations in accordance with the strength or degree of combination. This degree starts from two (i.e., d=2, where d is the degree of combinations).

Considering all combinations in a minimized test suite is a hard computational optimization problem [2, 10-12], because searching for the optimal set is an NP-hard problem [2, 11-15]. Hence, searching for an optimum set of test cases can be a difficult task, and finding a unified strategy that generates optimum results is challenging. Three approaches, namely, computational algorithms, mathematical construction, and nature-inspired metaheuristic algorithms, can be used to solve this problem efficiently and find a near-optimal solution [16].

Using nature-inspired metaheuristic algorithms can generate more efficient results than other approaches [17, 18]. This approach is more flexible than others because it can construct combinatorial sets with different input factors and levels. Hence, its outcome is more applicable because most real-world systems have different input factors and levels. Techniques that have been used to construct combinatorial sets include simulated annealing (SA) [7], tabu search (TS) [19], genetic algorithm (GA) [20], ant colony algorithm (ACA) [20, 21], and particle swarm optimization (PSO) [22, 23].

SA generates promising results in cases with small parameters and values as well as a small combination degree. However, it could not exceed certain parameters and values, and is unable to obtain results for combination degrees greater than three [20, 24]. PSO can compete with other strategies in most cases even when the combination degree exceeds three [25, 26]. However, PSO suffers from the effect of parameter tuning on its performance and from problems with local minima. Recent studies have discovered new nature-inspired metaheuristic algorithms that can produce better results than the traditional PSO algorithm for different applications.

Cuckoo Search (CS) [27] is one of the novel nature-inspired algorithms that have been proposed recently to solve complex optimization problems. CS can be used to efficiently solve global optimization problems [28] as well as NP-hard problems that cannot be solved by exact solution methods [29]. The most powerful feature of CS is its use of Lévy flights to update the search space for generating new candidate solutions. This mechanism allows the candidate solutions to be modified by applying many small changes during the iteration of the algorithm. This in turn makes a compromised relationship between exploration and exploitation which enhance the search capability [30]. To this end, recent studies proved that CS is potentially far more efficient than GA and PSO [31]. Such feature have motivated the use of CS to solve different kinds engineering problems such as scheduling problems [32], distribution networks [33], thermodynamics [34], and steel frame design [35].

The current paper presents the design and implementation of a strategy to construct optimized combinatorial sets using CS. Besides the Lévy flights, another advantage of CS over other counterpart nature-inspired algorithms such as PSO and GA, is that it does not have many parameters for tuning. Evidences showed that the generated results were independent of the value of the tuning parameters [27, 31].

The rest of the paper is organized as follows: Section 2 presents the mathematical notations, definitions, and theories behind the combinatorial testing. Section 3 illustrates a practical model of the problem using a real-world case study. Section 4 summarizes recent related works and reviews in the existing literature. Section 5 discusses the methodology of the research and implementation. The section reviews CS in detail and discusses the design and implementation of the strategy. In addition, it shows how the combinations are generated and describes in detail the algorithms that are used within the proposed strategy. Section 6 contains the evaluation results on the efficiency, performance, and effectiveness of CS. Section 7 presents threats to validity for the experiments and the case study. Finally, Section 8 concludes the paper.

## 2. Covering array mathematical preliminaries and notations

One future move toward combinatorial testing involves the use of a sampling strategy derived from a mathematical object called covering array (CA) [36]. In combinatorial testing, CA can be simply demonstrated by a table with rows and columns that contain the designed test cases; each row is a test case, and each column is an input factor for the software-under-test.

This mathematical object originates essentially from another object called orthogonal array (OA) [12]. An orthogonal array $OA_\lambda(N; d, k, v)$ is an $N \times k$ array in which for every $N \times d$ sub-array, each $d$-tuple occurs exactly $\lambda$ times, where $\lambda = N/v^d$. In this equation, d is the combination strength; k is the number of factors ($k \geq d$), and v is the number of symbols or levels associated with each factor. To consider all combinations, each d-tuple must occur at least once in the final test suite [37]. When each d-tuple occurs exactly one time, then $\lambda=1$, and it can be excluded from the mathematical notation, i.e., OA (N; d, k, v). As an example, the orthogonal array OA(9; 2, 4, 3) that contains three levels of value (v), with a combination degree (d) of two, and four factors (k) can be generated by nine rows. Figure 1(a) illustrates the arrangement of this array.

| OA (9; 2, 4, 3) | | | | CA (9; 2, 4, 3) | | | | MCA (9; 2, 4, $3^2 2^2$) | | | |
|---|---|---|---|---|---|---|---|---|---|---|---|
| $k_1$ | $k_2$ | $k_3$ | $k_4$ | $k_1$ | $k_2$ | $k_3$ | $k_4$ | $k_1$ | $k_2$ | $k_3$ | $k_4$ |
| 1 | 1 | 1 | 1 | 1 | 3 | 3 | 3 | 2 | 1 | 1 | 2 |
| 2 | 2 | 2 | 1 | 3 | 2 | 3 | 1 | 2 | 2 | 2 | 1 |
| 3 | 3 | 3 | 1 | 1 | 1 | 2 | 1 | 3 | 3 | 2 | 2 |
| 1 | 2 | 3 | 2 | 1 | 2 | 1 | 2 | 1 | 3 | 1 | 1 |
| 2 | 3 | 1 | 2 | 3 | 1 | 1 | 3 | 1 | 1 | 2 | 1 |
| 3 | 1 | 2 | 2 | 2 | 1 | 3 | 2 | 1 | 2 | 1 | 2 |
| 1 | 3 | 2 | 3 | 3 | 3 | 2 | 2 | 3 | 2 | 1 | 1 |
| 2 | 1 | 3 | 3 | 2 | 3 | 1 | 1 | 3 | 1 | 1 | 1 |
| 3 | 2 | 1 | 3 | 2 | 2 | 2 | 3 | 2 | 3 | 1 | 2 |
| (a) | | | | (b) | | | | (c) | | | |

Figure 1. Examples illustrating OA, CA, and MCA

However, the application of OA is limited by its requirement for uniform factors and levels; thus, this array is suitable for small test suites only [38, 39]. To address this limitation, the CA has been introduced to complement OA.

CA is another mathematical notation that is more flexible for representing large test suites with different parameters and values. In general, CA uses the mathematical expression $CA_\lambda$ (N; d, k, v) [1]. A covering array $CA_\lambda$(N; d, k, v) is an N × k array over $\{0, \ldots, v - 1\}$ such that every B $\in \binom{\{0,\ldots,k-1\}^d}{d}$ is $\lambda$-covered such that every N×d sub-array contains all ordered subsets from v values of size d at least λ times [40]. To consider all combinations, d-tuples must occur at least once. As such, we consider the value of λ=1, which is often omitted. Hence, the notation becomes CA(N; d, k, v) [41]. We say that the array has size N, combination degree d, k factors, v levels, and index λ. Given d, k, v, and λ, we denote the smallest N for which a $CA_\lambda$(N; t, k, v) exists as $CAN_\lambda$(d, k, g). A $CA_\lambda$(N ; d, k, v) with N = $CAN_\lambda$(d, k, v) is optimal as shown in Eq.1 [42]. Figure 1(b) shows a CA with N = 9, k = 4, v = 3, and d = 2.

$$CAN\ (d,k,v\ ) = min\{N : \exists\ CA\ (N, d, k, v\ )\} \ldots\ldots\ldots\ldots\ldots\ldots(1)$$

CA is suitable when the number of levels *v* is the same for each factor in the array. When factors have different numbers of levels, mixed covering array (MCA) is used. MCA is notated as MCA (*N, d, k*, ($v_1, v_2, v_3, \ldots, v_k$)). MCA is an *N* ×*k* array on *v* levels and *k* factors, where the rows of each *N* ×*d* sub-array cover all *d*-tuples of values from the *d* columns at least once [8]. For additional flexibility in the notation, the array can be presented by MCA (*N;d, $v^k$*)) and can be used for a fixed-level CA, such as CA (*N;d, $v^k$*) [14]. Figure 1(c) shows an MCA with size 9 that has four factors; two of these factors each have three levels, and the other two factors each have two levels, and each of these levels have two values.

## 3. Real-world problem model

Mozilla Firefox is a practical example that illustrates and models the concepts of combinatorial testing. Mozilla Firefox is a well-known Web browser that has many options and configurations that the user can control without difficulty because of its graphical user interface (GUI). Figure 2 shows a subset configuration of Mozilla Firefox, when many options of the scheme are combined to create a specific configuration. Configurations exist under various forms that enable them to be controlled in different ways, such as by clicking on the box or checking or unchecking an option. Users can change the configurations by clicking commands while operating Mozilla Firefox. Figure 2 shows a dialog box that contains six different configurations(i.e., warning when closing multiple tabs and warning when opening these tabs makes the browser operates lowly), with each configuration having two possible values (i.e., check and uncheck).The user can change the configuration based on the requirements.

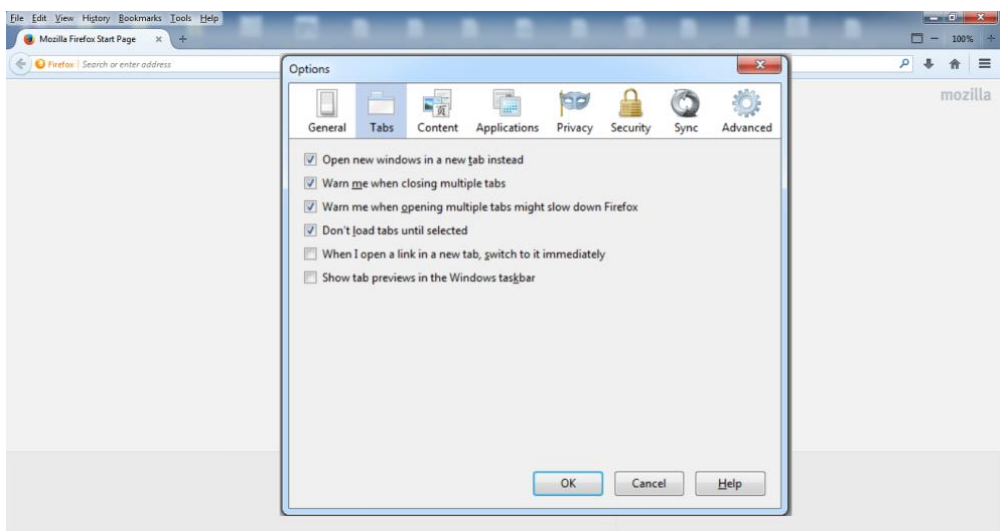

Figure 2. Subset configuration of Mozilla Firefox.

Testing the program by applying a set of designed test cases may reveal a set of different faults. However, evidence shows that applying the same set of test cases but with different configurations may lead to different faults [43, 44], which in turn leads us to consider different configurations for the same software-under-test. In addition, evidence shows that considering the interaction between the configurations (i.e., combination of configurations) will also detect new faults [26].

We need to consider that all the configurations must contain all possible combinations to test the software shown in Figure 2. Thus, the software has $2^6$ configurations, that is, 64 test cases. Xiao Qu called this collection of all possible combinations of configurations *configuration definition layer* (CDL) [43].Thus, a specific system that contains different configurations must be tested against its CDL, which leads to a configuration–aware testing process. Figure 3 shows this process.

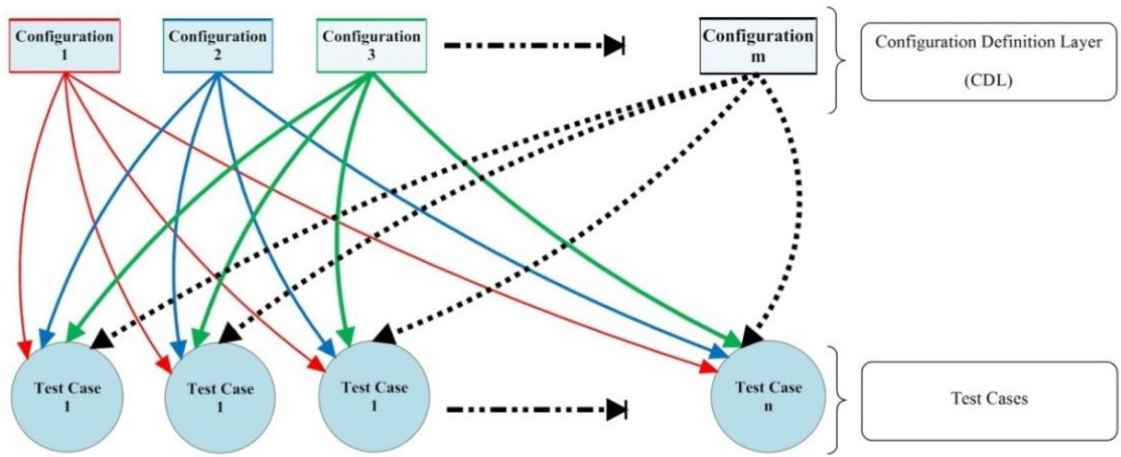

Figure 3. Relationship between test cases and configurations in configuration-aware testing

Ideally, each test case must be run against each configuration of the system. However, for large configurable software systems, considering all configurations is practically impossible because of time and resource constraints. For example, the command language interpreter of the Linux operating system (Bash) has approximately $7.6 \times 10^{23}$ possible configurations [44]. Reducing these configurations will dramatically minimize the time and cost of the testing process.

A sampling technique is needed to minimize these configurations systematically. Different sampling techniques are proposed in the literature (see [18, 45]). Among those techniques, combinatorial optimization effectively minimizes the number of configurations to be considered based on the combination degree. Combinatorial optimization can also be used to minimize the number of test cases. The final test suite can be represented mathematically by CA notation. The example in Figure 2 has six factors, each of which has two configurations. Considering combination degree $d$=2, the configuration set can be minimized to six configurations, that is, CA (6;2,6,2), by covering all the combinations of two configurations. However, CA (13;3,6,2), CA (26;4,6,2), and CA (33;5,6,2) represent configuration sets for combination degrees of 3, 4, and 5, respectively. Thus, instead of selecting all the combinations exhaustively, equivalence sets could lead to improved results with minimized time and cost.

## 4. Review of the literature and related works

As mentioned previously, generating a CA is an NP-hard problem .Thus, better methods have been sought. From the literature and other evidence, the generation methods have been confined to four main directions, namely, random method, mathematic method, greedy algorithm, and heuristic search algorithm [18].

Random methods are akin to ad hoc generation methods. In most cases, random methods work through a mechanism of random selection of a row of CA and by verifying whether it covers most of the combinations. The method continues to iterate until all the combinations are covered. This method is often used to show the effectiveness of other generation algorithms or to compare its fault detection abilities with other proposed methods [46, 47]. Although it obtains better results in some cases, random generation methods usually fail to achieve substantial results.[48].

The problem of CA generation has also been solved with extensions of OA construction that involve mathematical functions, regardless of the functions used for construction. Other mathematical methods use a recursive construction approach by building larger CAs from smaller CAs [41, 49]. Specifically, different tools use mathematical methods for construction, such as Combinatorial Test Services[50] and TConfig [51]. Although mathematical methods can effectively generate small-sized CAs, they fail to generate CAs for large parameters and values, particularly when the values are unequal among the parameters (i.e., MCA). These drawbacks limit its application for different cases of CA construction. However, the mathematical approach has the advantage of lightweight computation, which means it has a relatively fast generation time. In addition, the mathematical approach can produce optimal CAs for some special cases [40, 52, 53].

In addition to the aforementioned approaches, greedy algorithms and mathematical search methods solve the problem of CA generation computationally. Greedy algorithms are used to cover many uncovered combinations in each row of the CA. In this study, the CA rows are generated by using either of two methods, namely, one-row-at-a-time or one-parameter-at-a-time [18]. In the one-row-at-a-time method, the CA is constructed row by row. When a row is added, this row should essentially cover all $d$-tuples as much as possible. The construction process will stop when all $d$-tuples are covered successfully. The automatic efficient test generator (AETG) [54] is probably the first strategy that adopt this method of generation. The AETG strategy selects greedily one test case among several candidate test cases for each cycle. This algorithm serves as a base for a number of variations that have been developed later for AETG such as mAETG_SAT [7] and mAETG [24]. In addition to AETG, more work has been conducted on developing different algorithms and tools, such as the algorithm used for pairwise generation in the CATS tool [55], the greedy algorithms used in the Pairwise Independent Combinatorial Testing (PICT) tool [56], and the density-based greedy algorithm [57]. Most recently, pseudo-Boolean optimization is used with an AETG-like algorithm to generate efficient test suites [58]. Here, the strategy tries to do not reach maximum coverage of the d-tuples by the test cases. Instead, it tries to reach a balance point for the coverage ratio between [0.8, 0.9].

The one-parameter-at-a-time method attempts to construct rows of the generated CA by adding one parameter to each row each time and verifying the coverage of $d$-tuples periodically. Based on the coverage, parameters are added to the rows horizontally and vertically using heuristics until the CA is completed. The in-parameter-order (IPO) algorithm [10] was the pioneering implementation of this method. This strategy was further developed to produce variations of the IPO algorithm, such as IPOG [59], IPOG-D [14], IPOG-F [60], and IPO-s [61].

Heuristic search and artificial intelligence (AI)-based techniques have been applied effectively for CA construction. In general, these techniques start with a random set of solutions. Then, a transformation mechanism is applied to this set such that it is transferred to a new set in which its solutions are more efficient for $d$-tuple coverage. For each iteration, the transformation equations essentially create a more efficient set. Despite detailed variations in the heuristic search techniques, they essentially differ in transformation functions and mechanisms. Here, techniques, such as SA [7], TS[19], GA[20], ACA[20, 21], and PSO[22, 23], were used effectively for CA construction.

From a practical point of view, most of the time, the input factors of the real world applications suffer from the intertwined dependencies among each other which can potentially lead to problem in executing the test cases and may lead to failure due to improper execution [62]. Here, some of the parameters combinations are considered as impossible combination. Hence, they are considered as constraints in which they must be part of the final test suite. To this end, some of the recent strategies and tools start to support this issue such as mAETG, mAETG_SAT, IPOG, IPOG-D, SA, PICT, and TVG. However, strategies like GA, ACA, and PSO) generally do not show any evidence to support constraints. Recently, Garvin, B., et at. improve the SA algorithm to support constrained interaction testing [63]. To add the support for constraints, it is required to remove those combinations from the d-tuples list and add them directly to the final test suite.

Evidence showed that the computational methods (i.e., greedy and heuristic search algorithms) generate better results in terms of size. However, the computational methods may require more computational time than mathematical and random methods. In addition, the computational approach is more flexible than the other approaches as it can construct CAs with different parameters and values. Thus, the outcome of computational methods is more applicable because most real-world systems have different parameters and values rather than equal parameter values. Nonetheless, mathematical methods are useful for generating the optimal construction of CA in cases with few parameters and values, and a low degree of combination. As a result, computational methods are more applicable and more realistic, although they may not consistently produce the optimal CA.

As mentioned previously, different metaheuristic and AI-based strategies are proposed in the literature. Given an NP-hard problem, deriving a strategy that can generate optimal test cases for all parameters and values is practically impossible. To this end, researchers have attempted to construct better CAs in terms of size for most cases and to overcome the drawbacks of each method. In the case of small parameters and values, as well as a small combination degree $d$, SA usually generates promising results most of the time. However, SA is less effective when $d>3$. GA, ACA, and TS have also been applied in previous studies for generation [19, 20].By contrast, PSO can compete with other strategies when $d>3$ [25, 26]. However, PSO suffers from problems, such as parameter tuning, sticking in the local minima, and premature convergence of swarm problems which affect its optimization capability.

Particularly, these strategies suffer from heavy computation and inaccurate results for combinatorial test suite generation. For example, GA suffers from the crossover and mutation processes, which lead to heavy computation, ACA suffers from different problems when the number of ants increases, and TS suffers from the update mechanism of the tabu list sets. In addition, these strategies often impose a trade-off between reliability and speed of computation[34]. Today, no metaheuristic strategy can generate optimized results for all configurations, thereby implying that the investigation of new and efficient strategies with the help of metaheuristics is still an active research topic.

Cuckoo search (CS) has recently been found to be effective in solving engineering and optimization applications, with promising results. The convergence characteristics and results of CS are better than those of other metaheuristic optimization methods [28, 64]. In the literature, no studies have applied this promising method to generate combinatorial test suites. Thus, in this paper, we attempt to modify and apply the relative strengths of CS to this important part of software testing.

## 5. Cuckoo Search for Combinatorial Testing

Generating effective test cases and configurations is the most challenging task. As mentioned previously, testing the application exhaustively (i.e., test every possible event) is impossible most of the time because of time and resource constrains. Thus, an optimization strategy is needed to optimize and generate an optimized test suite that has the effectiveness of exhaustive testing. In this study, we use CS to search for test cases that cover all possible combinations at least once.

In this section, we provide the necessary details for the developed strategy. Section 5.1 presents the necessary background and illustrates the essential details of CS and its mechanism. Section 5.2 presents the details of the "all-combination-list generation" algorithm. Then, Section 5.3 presents the CS used for combinatorial testing and its optimization process and implementation.

### 5.1. *Cuckoo Search (CS)*

CS is a new metaheuristic search algorithm that was developed by Yang and Deb [27]. CS is inspired by the behavior of a fascinating bird called the cuckoo. The aggressive reproduction strategy of this bird inspired the researchers to study and investigate the opportunity to use its behavior within an optimization mechanism. Cuckoos lay their eggs in communal nests, although they may remove the eggs of another bird to increase the hatching probability of their own eggs. If the host bird discovers the eggs of the cuckoo, then it may throw the eggs away from the nest or may completely abandon the nest. The physiology and behavior of the cuckoo have the capability to mimic the appearance of the egg of the host.

The rules of the CS are as follows: (1) Each cuckoo selects a nest randomly to lay one egg in it, in which the egg represents a solution in a set of solutions.(2) Part of the nest contains the best solutions (eggs) that will survive to the next generation.(3) The probability of the host bird finding the alien egg in a fixed number of nests is $p_a \in [0,1]$[65]. If the host bird discovers the alien egg with this probability, then the bird will either discard the egg or abandon the nest to build a new one. Thus, we assumed that a part of $p_a$ with $n$ nest is replaced by new nests. Figure 4 shows the pseudocode and steps of the algorithm.

---

**Algorithm 1: Cuckoo Search**

1 Initialize a population of n host nests $x_i$, $i = 1, 2, \ldots, n$
2 **for** *all $x_i$* **do**
3     Calculate fitness $F_i = f(x_i)$
4 **end**
5 **while** *(Number of iterations <Max Number of iterations)*
6     *or (Stopping criteria satisfied)* **do**
7     Generate a cuckoo egg ($x_j$) by taking a Lévy flight from random nest
8     $F_j = f(x_j)$
9     Choose a random nest $i$
10     **if** $F_i > F_j$ **then**
11        $x_i \leftarrow x_j$
12        $F_i \leftarrow F_j$
13     **end**
14     Abandon a fraction pa of the worst nests
15     Build new nests at new locations via Lévy flights to replace nests lost
16     Evaluate fitness of new nests and rank all solutions
17 **end**

---

[66]

Figure 4. Pseudo code of Cuckoo Search [27]

Lévy flight is used in the cuckoo algorithm to conduct local and global searches [67]. Here, Lévy flight serves as an update mechanism to update and modify the initial random search space. This update mechanism allows the algorithm to generate new candidate solutions by applying small changes during the iteration which behave like a step toward the best solution [30]. The rule of Lévy flight is used successfully in stochastic simulations of different applications, such as biology and physics. Lévy flight is a random path of walking that takes a sequence of jumps, which are selected from a probability function. A step can be represented by the following equation for the solution $x^{(t+1)}$ of cuckoo $i$:

$$x_i^{t+1} = x_i^{(t)} + \alpha \oplus \text{Lévy}(\lambda) \quad \text{................................ (2)}$$

where $\alpha$ the size of each step in which $\alpha > 0$ and depends on the optimization problem scale. The product $\oplus$ is the entrywise multiplication, and Lévy ($\lambda$) is the Lévy distribution. The algorithm continues to move the eggs to another position if the objective function found better positions.

Another advantage of CS over other counterpart stochastic optimization algorithms, such as PSO and GA, is that it does not have many parameters for tuning. The only parameter for tuning is $p_a$. Yang and Deb [27, 31] obtained evidence from the literature and showed that the generated results were independent of the value of this parameter and can be fit to a proposed value $p_a$=0.25.

## 5.2. *The d-tuples list generation algorithm*

Generating the *d*-tuples list is essential to calculate the fitness function $F_i = f(x_i)$. The *d*-tuples list contains all possibilities of combinations between input factors *k*. As an example, we consider a system with three input factors ($k_1 k_2 k_3$), each factor having three levels ($v_1\ v_2\ v_3$). For exhaustive testing, when the combination degree *d*=3 (i.e., *d*=*k*), (3×3×3) combinations result in 9 combinations. However, as mentioned previously, exhaustive testing is impossible. Thus, lower combination degrees are considered to minimize the test cases. For example, when *d*=2, the combinations are ($k_1 k_2$), ($k_1 k_3$), and ($k_2 k_3$). In turn, these combinations are converted to the all-combination-list, which contains (3×3) + (3×3) + (3×3) = 27 combinations with *d*=2. Then, this list will be covered row by row during the optimization process.
Generation this list is difficult because of its tightness with the combination degree. Thus, the generation of the list starts by considering the number of factors and then calculating the binary equivalence numbers of ($2^k - 1$). This algorithm is implemented in the "Generate Binary Digits" function, as shown in Figure 5.

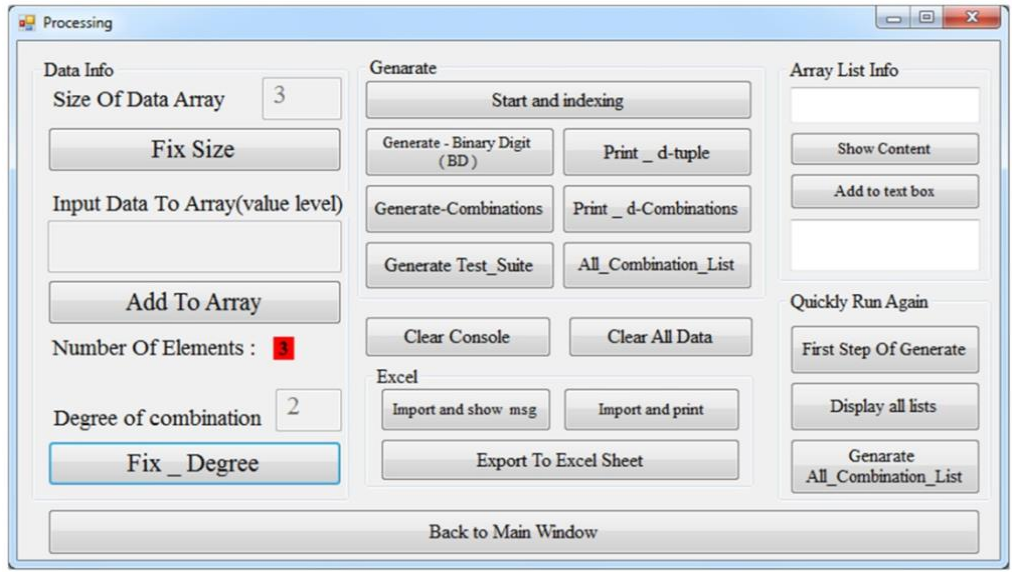

Figure 5. Main window of the implemented strategy

The algorithm starts by inputting binary digits from 0 to ($2^k - 1$) in a list. For example, when *k*=3, then the list contains (000) to (111). A filtering mechanism is combined with the algorithm to filter the number of (1's) in each number from the list depending on the combination degree. For example, when *d*=2, then the binary numbers after filtering are [(011), (101), (110)], which are equivalent to [($k_2 k_3$), ($k_1 k_3$), ($k_1 k_2$),], which, in turn, serves as a master algorithm for generating combinations of input factors for all degrees. The progress and output of this algorithm can be noted clearly in the output screen of the strategy shown in Figure 6.

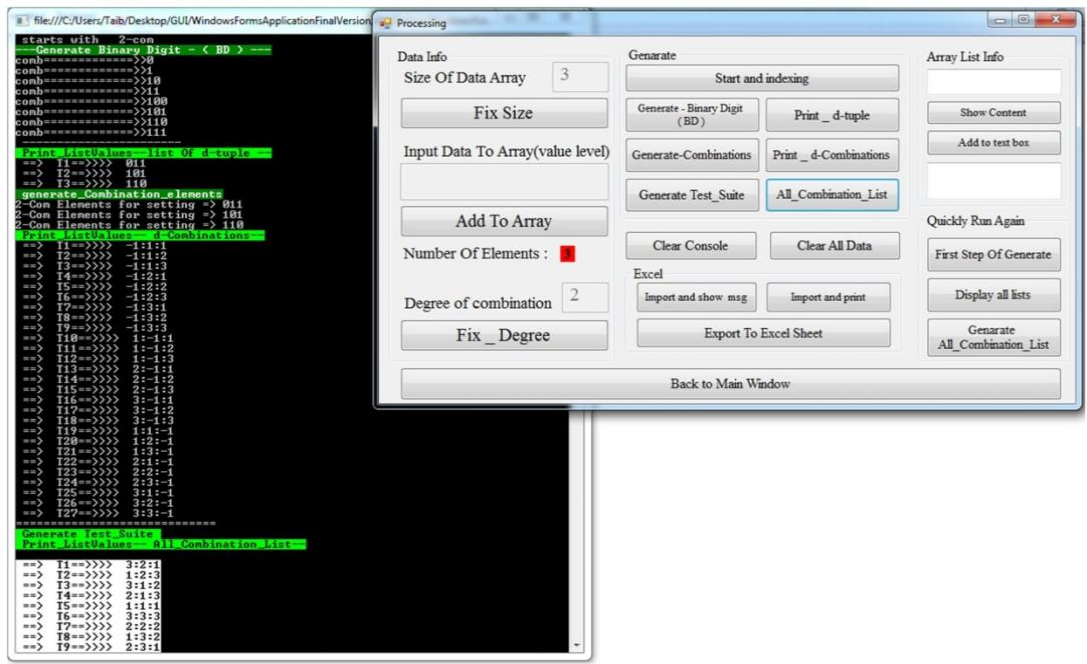

Figure 6. Strategy in progress when each algorithm is executed and the final optimized set is generated.

When the combinations of factors are identified, the values of the corresponding factor are matched. This algorithm is implemented in the "Generate-Combinations" function, as shown in Figures 5 and 6. When a factor is missed in the combination (i.e., its corresponding binary value is 0), its corresponding value will be "don't care" because we do not need its value for that specific combination (in this study, −1 is used as an indication only). Upon completion of this algorithm, all the combinations are stored in a list to be used for calculating the fitness function of the CS. The output list of this algorithm can be noted clearly in the output screen of the strategy shown in Figure 6.

An algorithm is used to assess the search process for the combinations efficiently. In this study, the rows in the $d$-tuples list are stored in groups. Each group is assigned an index number that indicates its position in the list. The groups are selected based on the combination of factors. For example, in the aforementioned sample, the combination ($k_2k_3$) is stored in the index from (0 to 8) because it has nine rows of combinations. Thus, the next group is stored in the index from (9 to 17).

### 5.3. *Optimization process with Cuckoo Search*

When the $d$-tuples list is generated, then CS starts. In this study, the CS algorithm is modified to solve the current problem. The fitness function is used to derive the better solution among a set of solutions. In this study, a row with higher fitness weight is defined as a row that can cover a higher number of rows in the $d$-tuples list. Figure 7 shows the pseudocode of combinatorial test suite generation in which the CS is modified for this purpose.

```
Algorithm 2: Combinatorial test suite generation
   Input: Input-factors k and levels v
   Output: A test case
 1 Let d-tuples list be a set of all combinations' list that must be covered
 2 Initialize a population of m host nests x_i, i = 1, 2, ..., m
 3 for all x_i do
 4     Calculate the coverage of combinations and return the weight
 5 end
 6 Iteration number Iter ← 1
 7 while (Number of iterations <Max Number of iterations)
 8     or (d-tuples list is not empty) do
 9     Iter ← Iter + 1
10     Sort the nest by the weight of combination's coverage
11     for all nests to be abandoned do
12         Current position x_i
13         Perform Lévy flight from x_i to generate new egg x_j
14         x_i ← x_j
15         F_i ← f(x_i)
16     end
17     for all of the top nests do
18         Current position x_i
19         perform Lévy flight from x_i to generate new eggs x_k
20         F_k ← f(x_k)
21         if F_k >F_i then
22             x_i ← x_k
23             F_i ← F_k
24         end
25     end
26 end
27 Add the first nest to the final test suite
28 Remove all the related combinations from the d-tuples list
```

Figure 7. Pseudocode of combinatorial test case generation with CS

As shown in Figure 7, the strategy starts by considering the input configuration. Then, the *d*-tuples list is generated. CS starts by initializing a random population that contains a number of nests. Given that the number of levels for each input factor is a discrete number, the initialized population is discrete, not an open interval. Thus, the population is initialized with a fixed interval between 0 and $v_i$. In this study, a system has different factors in which a test case is a composite of more than two factors that form a row in the final test suite. As a result of such an arrangement, each test case is treated as a vector $x_i$ that has dimensions equal to the number of input factors of the system. In addition, the levels for each input factor are basically an integer value. As a result, each dimension in the vector-initialized population must be an integer value.

Although the initial population is initialized in a discrete interval, the algorithm can produce out-of-the-bound levels for the input factors. Thus, the vector must be restricted with lower and upper bounds. The rationale behind this restriction is that the cuckoo lays its eggs in the nests that are recognized by its eyes.

When the CS iterates, it uses Lévy flight to walk toward the optimum solution. Lévy flight is a walk that uses random steps in which the length of each step is determined by Lévy distribution. The generation of random steps in Lévy flight consists of two steps [68], namely, the generation of steps and the choice of random direction. The generation of direction normally follows a uniform distribution. However, in the literature, the generation of steps follows a few methods. In this study, we follow the Mantegna algorithm, which is the most efficient and effective step generation method. Within this algorithm, a step length *s* can be defined as follows:

$$s = \frac{u}{|v|^{1/\beta}} \ldots\ldots\ldots\ldots\ldots\ldots\ldots (3)$$

where *u* and *v* are derived from the normal distribution in which

$$u \sim N(0, \sigma_u^2) \qquad v \sim N(0, \sigma_u^2) \ldots\ldots\ldots\ldots\ldots (4)$$

$$\sigma_u = \left\{ \frac{\Gamma(1+\beta)\sin\left(\frac{\pi\beta}{2}\right)}{\Gamma\left[\frac{1+\beta}{2}\right]\beta 2^{\frac{\beta-1}{2}}} \right\}^{1/\beta}, \qquad \sigma_v = 1 \ldots\ldots\ldots\ldots\ldots (5)$$

Based on the aforementioned design constraints, the complete strategy steps, including the CS, are summarized in Figure 7.

As mentioned previously, the strategy starts by considering the input configuration. Normally, the input is a composite input with the factors, levels, and desired combination degree *d*. The combination degree *d*>1 and is less than the number of input factors. Using the d-tuples generation algorithm described previously, the *d*-tuples list is generated (Step 1). Then, the strategy uses the CS, which starts by initializing a population with *m* nests, with each nest consisting of dimensional vectors equal to the number of factors that have a number of levels (Step 2). From a practical point of view, each nest contains a candidate test case for the final test suite. Then, the CS starts to assess each nest by evaluating coverage capability of the *d*-tuples (Steps 3-5). This mechanism is used to assess the fitness function of the CS. The fitness function *f(x_i)* of the test case $x_i$ in this strategy is defined as:

$$f(x_i) = \sum_{i \in new\ d-tuple(x_i)}^{i} d_i \ldots\ldots\ldots\ldots (6)$$

where *d-tuples($x_i$)* indicates new tuples that are not covered by the previous generated tests but covered by the test $x_i$. $d_i$ denotes the strength of the interaction *i*. For example, when a nest can cover four d-tuples, then its weight of coverage is 4. The strategy uses a special mechanism described previously (Section 5.2) to determine the number of covered tuples and to verify the weight. Based on the results of coverage for all nests, the strategy sorts the nests again in the search space based on the highest coverage (Step 10). The lowest coverage in the search space will be abandoned. For the abandoned nests, a Lévy flight is conducted to verify the availability of better coverage (Step 11). If better coverage is obtained for a specific nest, then the nest is replaced by the current nest content (Steps 12-15). This process serves just like global search in other optimization algorithms. Then, for all of the top nests after sorting, a Lévy flight is conducted to search for the local best nests (Steps 17-19). If better coverage is obtained after the Lévy flight for a specific nest, then the nest is replaced with the one that have better coverage (Steps 20-24). These steps (Steps 9 to 26) in the CS will update the search space for each iteration.

Two stopping criteria are defined for the CS. First, if the nest reaches the maximum coverage, then the loop will stop and the algorithm will add this test case to the final test suite and remove its related tuples in the *n*-tuples list. Second, if the *d*-tuples list is empty, then no combinations are covered. If the iteration reaches the final iteration, then the algorithm will select the best coverage nest to be added to the final test suite (Step 27) and remove the related tuples in the *n*-tuples list (Step 28). Figure 8 shows a graphical representation of the strategy to summarize the aforementioned steps for better understanding. The sequence of running is show in red circle on the figure.

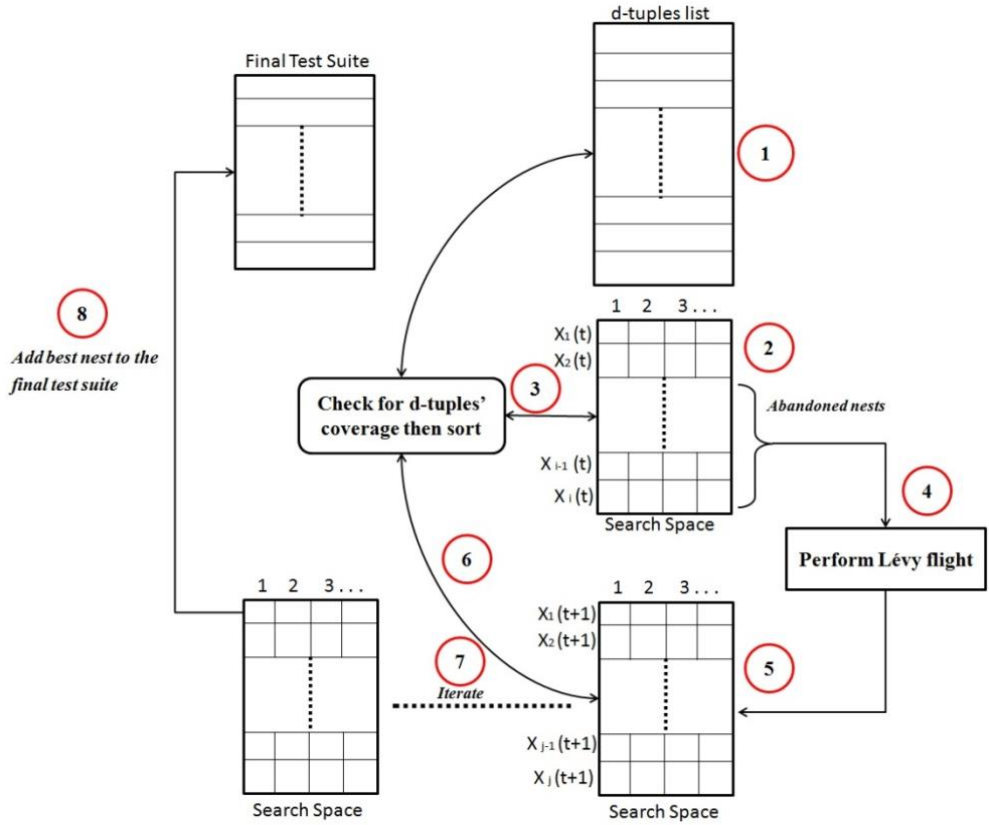

Figure 8. Graphical representation of CS strategy

The constructed final test suite can be noted clearly in Figure 6. This mechanism will continue as far as *n*-tuples remain in the list. Figure 9 shows a running example to illustrate how the tuples are covered and removed and how the final test suite is constructed.

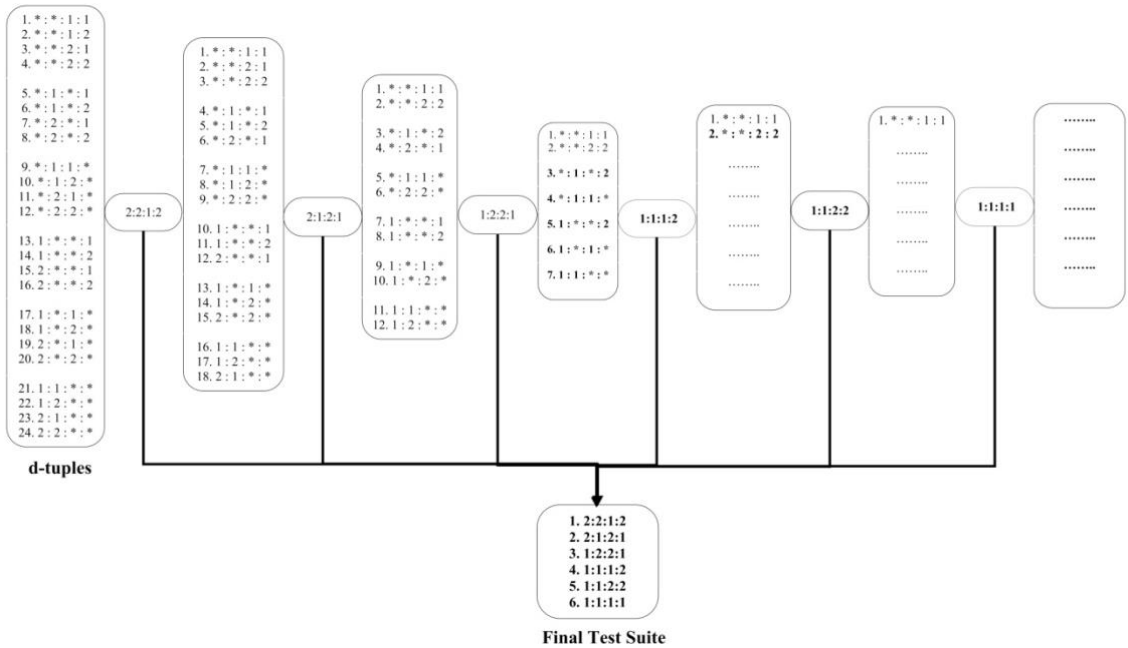

Figure 9: Running example to show the construction of final test suite and the removal of *d*-tuples CA(*N*;2, $2^4$).

## 6. Evaluation results and discussion

The evaluation phase for the proposed strategy is divided into the following three main sections: (1) evaluation of the generation efficiency,(2) evaluation of the generation performance, and (3) evaluation of the effectiveness of the generated test suite. Based on the literature [16, 26, 69], efficiency is evaluated based on the size of the generated test suite, whereas performance is evaluated based on the time taken by the strategy to generate a specific test suite. For these two evaluation phases, the strategy is compared with other available strategies.

Some strategies are available publicly as tools to be downloaded and installed. Other strategies are unavailable publicly, yet their evaluations are published for certain cases. We consider the performance evaluation for strategies that are available for implementation within the same evaluation environment. By contrast, for unavailable strategies, we consider the efficiency evaluation only. The rationale behind this option is that the efficiency criterion is not affected by the research environment as the size of the CA is not affected by computer speed. However, installing all the tools in the same environment is essential to ensure a fair comparison of performance as the construction time is affected by the specifications of the computer.

The effectiveness of the generated test suite is evaluated by adopting a case study on a reliable artifact program to prove the applicability and correctness of the strategy for a real-world software testing problem. Given that the generated test suite did not consider the internal structure of the artifact program, the testing process represents a functional testing process that considers the program configuration.

The experimental environment consists of a desktop PC with Windows 7, 64-bit, 2.5 GHz, Intel Core i5 CPU, and 6 GB of RAM. The algorithms are coded and implemented in C#.

### 6.1. *Efficiency evaluation*

The efficiency of the combinatorial test suite construction is measured by the size of the test suite generated by the strategy. For strategies that depend on metaheuristic algorithms, a degree of randomness is observed, especially when the strategy starts with a random solution. As a result, the produced results are nondeterministic. For this reason, the published results in the literature were achieved by considering the best achieved size of either 5 or 10 runs [21, 24, 63, 70]. In this study, each experiment is conducted 40 times, and the best and average sizes are reported for each test to account for the statistical significance and to derive and report better representation of the performance.

The proposed strategy is compared with seven well-known strategies, namely, Jenny [71], TConfig [72], PICT[73], TVG [74], IPOG [59], IPOG-D [14], and PSO [26]. Some tools, such as PICT and IPOG, have options to make them behave randomly. However, the literature indicates that the deterministic results are more efficient than the random results. Hence, the deterministic option is used within these tools. Several experiments based on the experiments reported in the literature are established to determine the competitiveness. Different CA configurations are achieved by varying the degree of combination ($d$), number of components ($k$), and number of levels ($v$) to select and establish the experiments. These configurations enable the observer to determine the efficiency and performance of the proposed strategy in constructing different types of CA. Usually, the experiments are divided into two sets of experiments, as follows: (1) The strategy is compared (in terms of generated size $N$) with the deterministic computational strategies.(2) The strategy is compared (in terms of generated size $N$) with the published results of those strategies depending on metaheuristic algorithms. Tables 2 to 12 show the comparative results of the aforementioned cases. The best result for each configuration in each table is shown in bold numbers. Cells marked NA (not available) indicate that the results are unavailable from the literature, and cells marked NS (not supported) indicate that the strategy does not support the construction of that specific configuration. In Table 2, the strategies AETG, mAETG, GA, SA, and ACA are not publically available for download. Hence, the results are adopted from the experiments undertaken by [24] and [20]. For the rest of the tables, the strategies are downloaded and executed within our environment. For fair comparison, Table 1 shows the parameter tuning values of each meta-heuristic algorithm used in the comparison.

Table 1. Parameters used for the existing meta-heuristic algorithms

| Algorithm | Parameter | Values |
|---|---|---|
| **GA [20]** | Population size | 25 |
| | Best cloned | 1 |
| | Tournament selection | 0.8 |
| | Random crossover | 0.75 |
| | Gene mutation | 0.03 |
| | Max stale period | 3 |
| | Escape mutation | 0.25 |
| | Iteration | 1000 |
| **SA [75]** | starting temperature | 20 |
| | cooling schedule | 0.9998 |
| | Iteration | 1000 |
| **ACA [20]** | Number of ants | 20 |
| | Pheromone control | 1.6 |
| | Heuristic control | 0.2 |
| | Pheromone persistence | 0.5 |
| | Initial pheromone | 0.4 |
| | Pheromone amount | 0.01 |
| | Elite ants | 2 |
| | Max stale period | 5 |
| | Iteration | 1000 |
| **PSO [26]** | Population size | 80 |
| | Acceleration coefficients | 1.375 |
| | Inertia weight | 0.3 |
| | Iteration | 100 |
| **CS** | Population size | 100 |
| | Probability $p_a$ | 0.25 |
| | Iteration | 100 |

Table 2. Comparison with existing meta-heuristic algorithms for different configuration

| | k=10 | | | | | | | |
|---|---|---|---|---|---|---|---|---|
| | AETG [20] | mAETG [24] | GA [20] | SA [20] | ACA [20] | PSO | CS | |
| | N | N | N | N | N | N | Best. N | Avg. N |
| CA(N; 2, $3^4$) | **9** | **9** | **9** | **9** | **9** | **9** | 9 | *9.8* |
| CA(N; 2, $3^{13}$) | **15** | 17 | 17 | 16 | 17 | 17 | 20 | *22.4* |
| MCA(N; 2, $5^1 3^8 2^2$) | 19 | 20 | **15** | **15** | 16 | 21 | 21 | *22.6* |
| MCA(N; 2, $6^1 5^1 4^6 3^8 2^3$) | 34 | 35 | 33 | **30** | 32 | 39 | 43 | *45.4* |
| MCA(N; 2, $7^1 6^1 5^1 4^6 3^8 2^3$) | 45 | 44 | **42** | **42** | **42** | 48 | 51 | *52.4* |
| CA(N; 3, $3^6$) | 47 | 38 | **33** | **33** | **33** | 42 | 43 | *44.8* |
| CA(N; 3, $4^6$) | 105 | 77 | **64** | **64** | **64** | 102 | 105 | *108.2* |
| CA(N; 3, $5^7$) | 229 | 218 | 218 | **201** | 218 | 229 | 233 | *236.2* |
| CA(N; 3, $6^6$) | 343 | 330 | 331 | **300** | 330 | 338 | 350 | *360.4* |
| MCA(N; 3, $10^1 6^2 4^3 3^1$) | NA | 377 | **360** | **360** | 361 | 385 | 393 | *399.8* |

Table 3. Size of variable input configurations when $3 \leq k \leq 12$, each having three levels and $2 \leq d \leq 6$.

| | | \multicolumn{9}{c}{$v = 3$} | | | | | | | | |
|---|---|---|---|---|---|---|---|---|---|---|
| | k | Jenny | TConfig | ITCH | PICT | TVG | CTE-XL | IPOG | PSO | \multicolumn{2}{c}{CS} |
| | | N | N | N | N | N | N | N | N | Best.N | AVG.N |
| d=2 | 3 | **9** | 10 | **9** | 10 | 10 | 10 | 11 | **9** | **9** | *9.6* |
| | 4 | 13 | 10 | **9** | 13 | 12 | 14 | 12 | **9** | **9** | *10.0* |
| | 5 | 14 | 14 | 15 | 13 | 13 | 14 | 14 | 12 | **11** | *11.8* |
| | 6 | 15 | 15 | 15 | 14 | 15 | 14 | 15 | **13** | **13** | *14.2* |
| | 7 | 16 | 15 | 15 | 16 | 15 | 16 | 17 | 15 | **14** | *15.6* |
| | 8 | 17 | 17 | **15** | 16 | **15** | 17 | 17 | **15** | **15** | *15.8* |
| | 9 | 18 | 17 | **15** | 17 | **15** | 18 | 17 | 17 | 16 | *17.2* |
| | 10 | 19 | 17 | **15** | 18 | 16 | 18 | 20 | 17 | 17 | *17.8* |
| | 11 | 17 | 20 | **15** | 18 | 16 | 20 | 20 | 17 | 18 | *18.6* |
| | 12 | 19 | 20 | **15** | 19 | 16 | 20 | 20 | 18 | 18 | *18.8* |
| d=3 | 4 | 34 | 32 | **27** | 34 | 34 | 34 | 39 | 30 | 28 | *29.0* |
| | 5 | 40 | 40 | 45 | 43 | 41 | 43 | 43 | 39 | **38** | *39.2* |
| | 6 | 51 | 48 | 45 | 48 | 49 | 52 | 53 | 45 | **43** | *44.2* |
| | 7 | 51 | 55 | **45** | 51 | 55 | 54 | 57 | 50 | 48 | *50.4* |
| | 8 | 58 | 58 | **45** | 59 | 60 | 63 | 63 | 54 | 53 | *54.8* |
| | 9 | 62 | 64 | 75 | 63 | 64 | 66 | 65 | **58** | 58 | *59.8* |
| | 10 | 65 | 68 | 75 | 65 | 68 | 71 | 68 | **62** | **62** | *63.6* |
| | 11 | 65 | 72 | 75 | 70 | 69 | 76 | 76 | **64** | 66 | *68.2* |
| | 12 | 68 | 77 | 75 | 7 | 70 | 79 | 76 | **67** | 70 | *71.8* |
| d=4 | 5 | 109 | 97 | 153 | 100 | 105 | NA | 115 | 96 | **94** | *95.8* |
| | 6 | 140 | 141 | 153 | 142 | 139 | NA | 181 | 133 | **132** | *134.2* |
| | 7 | 169 | 166 | 216 | 168 | 172 | NA | 185 | 155 | **154** | *156.8* |
| | 8 | 187 | 190 | 216 | 189 | 192 | NA | 203 | 175 | **173** | *174.8* |
| | 9 | 206 | 213 | 306 | 211 | 215 | NA | 238 | **195** | 195 | *197.8* |
| | 10 | 221 | 235 | 336 | 231 | 233 | NA | 241 | **210** | 211 | *212.2* |
| | 11 | 236 | 258 | 348 | 249 | 250 | NA | 272 | **222** | 229 | *231.0* |
| | 12 | 252 | 272 | 372 | 269 | 268 | NA | 275 | **244** | 253 | *255.8* |
| d=5 | 6 | 348 | 305 | NA | 310 | 321 | NA | 393 | 312 | **304** | *307.8* |
| | 7 | 458 | 477 | NA | 452 | 462 | NA | 608 | 441 | **434** | *440.2* |
| | 8 | 548 | 583 | NA | 555 | 562 | NA | 634 | **515** | 515 | *517.8* |
| | 9 | 633 | 684 | NA | 637 | 660 | NA | 771 | 598 | **590** | *593.8* |
| | 10 | 714 | 773 | NA | 735 | 750 | NA | 784 | **667** | 682 | *688.0* |
| | 11 | 791 | 858 | NA | 822 | 833 | NA | 980 | **747** | 778 | *780.2* |
| | 12 | 850 | 938 | NA | 900 | 824 | NA | 980 | **809** | 880 | |
| d=6 | 7 | 1087 | **921** | NA | 1015 | 1024 | NA | 1281 | 977 | 963 | *970.8* |
| | 8 | 1466 | 1515 | NA | 1455 | 1484 | NA | 2098 | 1402 | **1401** | *1410.8* |
| | 9 | 1840 | 1931 | NA | 1818 | 1849 | NA | 2160 | **1648** | 1689 | *1695.4* |
| | 10 | 2160 | NA | NA | 2165 | 2192 | NA | 2726 | **1980** | 2027 | *2035.4* |
| | 11 | 2459 | NA | NA | 2496 | 2533 | NA | 2739 | **2255** | 2298 | *2302.2* |
| | 12 | 2757 | NA | NA | 2815 | 2597 | NA | 3649 | **2528** | 2638 | *2640.6* |

Table 4. Size of variable input configurations when 2 ≤ v ≤ 5, each having seven factors and 2 ≤ d ≤ 6

| | | | | | | | | | | CS | |
|---|---|---|---|---|---|---|---|---|---|---|---|
| | | Jenny | TConfig | ITCH | PICT | TVG | CTE-XL | IPOG | PSO | CS | |
| | $v$ | N | N | N | N | N | N | N | N | Best.N | AVG.N |
| | | | | | | $k=7$ | | | | | |
| $d=2$ | 2 | 8 | 7 | **6** | 7 | 7 | 8 | 8 | **6** | **6** | *6.8* |
| | 3 | 16 | **15** | **15** | 16 | **15** | 16 | 17 | **15** | **15** | *16.2* |
| | 4 | 28 | 28 | 28 | 27 | 27 | 30 | 28 | 26 | **25** | *26.4* |
| | 5 | **37** | 40 | 45 | 40 | 42 | 42 | 42 | **37** | **37** | *38.6* |
| $d=3$ | 2 | 14 | 16 | 13 | 15 | 15 | 15 | 19 | 13 | **12** | *13.8* |
| | 3 | 51 | 55 | **45** | 51 | 55 | 54 | 57 | 50 | 49 | *51.6* |
| | 4 | 124 | **112** | **112** | 124 | 134 | 135 | 208 | 116 | 117 | *118.4* |
| | 5 | 236 | 239 | 225 | 241 | 260 | 265 | 275 | 225 | **223** | *225.4* |
| $d=4$ | 2 | 31 | 36 | 40 | 32 | 31 | NA | 48 | 29 | **27** | *29.6* |
| | 3 | 169 | 166 | 216 | 168 | 167 | NA | 185 | **155** | **155** | *156.8* |
| | 4 | 517 | 568 | 704 | 529 | 559 | NA | 509 | **487** | **487** | *490.2* |
| | 5 | 1248 | 1320 | 1750 | 1279 | 1385 | NA | 1349 | 1176 | **1171** | *1175.2* |
| $d=5$ | 2 | 57 | 56 | NA | 57 | 59 | NA | 128 | **53** | **53** | *54.2* |
| | 3 | 458 | 477 | NA | 452 | 464 | NA | 608 | 441 | **439** | *442.2* |
| | 4 | 1938 | 1792 | NA | 1933 | 2010 | NA | 2560 | **1426** | 1845 | *1850.8* |
| | 5 | 5895 | NA | NA | 5814 | 6257 | NA | 8091 | **5474** | 5479 | *5485.2* |
| $d=6$ | 2 | 87 | **64** | NA | 72 | 78 | NA | **64** | **64** | 66 | *67.2* |
| | 3 | 1087 | **921** | NA | 1015 | 1016 | NA | 1281 | 977 | 973 | *978.4* |
| | 4 | 6127 | NA | NA | 5847 | 5978 | NA | 4096 | 5599 | 5610 | *5620.8* |
| | 5 | 23492 | NA | NA | 22502 | 23218 | NA | 28513 | **21595** | 21597 | *21610.8* |

Table 5. Size of variable input configurations when 2 ≤ v ≤ 6, each having ten factors and 2 ≤ d ≤ 6

| | | | | | | | | | CS | |
|---|---|---|---|---|---|---|---|---|---|---|
| | | Jenny | TConfig | ITCH | PICT | TVG | IPOG | PSO | CS | |
| | $v$ | N | N | N | N | N | N | N | Best.N | AVG.N |
| | | | | | $k=10$ | | | | | |
| $d=4$ | 2 | 39 | 45 | 58 | 43 | 40 | 49 | 34 | **28** | *30.4* |
| | 3 | 221 | 235 | 336 | 231 | 228 | 241 | 213 | **211** | *212.8* |
| | 4 | 703 | 718 | 704 | 742 | 782 | 707 | **685** | 698 | *701.8* |
| | 5 | 1719 | 1875 | 1750 | 1812 | 1917 | 1965 | **1716** | 1731 | *1740.2* |
| | 6 | **3519** | NA | NA | 3735 | 4195 | 3335 | 3880 | 3894 | *3902.6* |

Table 6. Size of variable input configurations when k=10, each having five levels when 2 ≤ d ≤ 6

| | | k = 10 | | | | | | | | | |
|---|---|---|---|---|---|---|---|---|---|---|---|
| | d | Jenny | TConfig | ITCH | PICT | TVG | CTE-XL | IPOG | PSO | CS | |
| | | N | N | N | N | N | N | N | N | Best.N | AVG.N |
| v=5 | 2 | **45** | 48 | **45** | 47 | 50 | 50 | 50 | **45** | 45 | *47.8* |
| | 3 | 290 | 312 | **225** | 310 | 342 | 347 | 313 | 287 | 297 | *299.2* |
| | 4 | 1719 | 1878 | 1750 | 1818 | 1971 | NS | 1965 | **1716** | 1731 | *1740.2* |
| | 5 | 9437 | NA | NS | 9706 | NA | NS | 11009 | **9425** | 9616 | *9620.4* |
| | 6 | NA | NA | NS | **47978** | NA | NS | 57290 | 50350 | 50489 | *50503.6* |

Table 7. Size of variable input configurations when k=10, each having two levels when 2 ≤ d ≤ 6

| | | k=10 | | | | | | | |
|---|---|---|---|---|---|---|---|---|---|
| | d | Jenny | TConfig | ITCH | TVG | IPOG | PSO | CS | |
| | | N | N | N | N | N | N | Best.N | AVG.N |
| v=2 | 2 | 10 | 9 | **6** | 10 | 10 | 8 | 8 | 9.0 |
| | 3 | 18 | 20 | 18 | 17 | 19 | 17 | **16** | 17.4 |
| | 4 | 39 | 45 | 58 | 41 | 49 | 37 | **36** | 38.2 |
| | 5 | 37 | 95 | NS | 84 | 128 | 82 | **79** | 81.8 |
| | 6 | 169 | 183 | NS | 168 | 352 | 158 | **157** | 160.2 |

Table 8. Size of variable input configurations when d=4, and 5 ≤ k ≤ 12 each having five levels

| | | v=5 | | | | | | | | |
|---|---|---|---|---|---|---|---|---|---|---|
| | k | Jenny | TConfig | ITCH | PICT | TVG | IPOG | PSO | CS | |
| | | N | N | N | N | N | N | N | Best.N | AVG.N |
| d=4 | 5 | 837 | 773 | **625** | 810 | 849 | 908 | 779 | 776 | 781.8 |
| | 6 | 1074 | 1092 | **625** | 1072 | 1128 | 1239 | 1001 | 991 | 1002.4 |
| | 7 | 1248 | 1320 | 1750 | 1279 | 1384 | 1349 | 1209 | **1200** | 1205.4 |
| | 8 | 1424 | 1532 | 1750 | 1468 | 1595 | 1792 | 1417 | **1415** | 1420.6 |
| | 9 | 1578 | 1724 | 1750 | 1643 | 1795 | 1793 | 1570 | **1562** | 1672.4 |
| | 10 | 1791 | 1878 | 1750 | 1812 | 1971 | 1965 | **1716** | 1731 | 1740.2 |
| | 11 | 1839 | 2038 | **1750** | 1957 | 2122 | 2091 | 1902 | 2062 | 2070.6 |
| | 12 | 1964 | NA | **1750** | 2103 | 2268 | 2258 | 2015 | 2223 | 2230.8 |

Table 9. TCAS module (MCA (N; d, $2^7 3^2 4^1 10^2$)

| | k=12 | | | | | | | |
|---|---|---|---|---|---|---|---|---|
| d | Jenny | TConfig | ITCH | TVG | IPOG | PSO | CS | |
| | N | N | N | N | N | N | Best.N | AVG.N |
| 2 | 108 | 108 | 120 | 101 | **100** | **100** | **100** | 104.2 |
| 3 | 413 | 472 | 2388 | 434 | **400** | 400 | 410 | 415.2 |
| 4 | 1536 | 1548 | 1484 | 1599 | **1377** | 1520 | 1537 | 1540.0 |
| 5 | 4621 | NS | NS | 4773 | **4283** | 4566 | 4566 | 4576.2 |
| 6 | **11625** | NS | NS | NS | 11939 | 11743 | 11431 | 11450.0 |

Table 2 shows the results of the comparison of the generated size for the test suites using our strategy and its counterpart metaheuristic algorithms (i.e., SA, GA, ACA, and PSO). As mentioned previously, these strategies are unavailable publicly. Thus, the results are directly compared with the results published in [20] and [22]. In addition, the results of AETG and mAETG strategies are considered in the comparison because the GA and ACA results are derived from them. The results of AETG and mAETG are adopted from [24]. The comparison is fair because the size of the generated test suite is unaffected by the specification and the environment of the implemented strategy.

The table shows the size of the smallest generated sizes for the combinatorial test suite when $2 \leq d \leq 3$. Notably, SA generates better results for most configurations. GA, PSO, ACA, and CS generate comparable results. However, in most cases, CS generates better results than PSO because of its optimization capability. Notably, although SA generates better results, it fails to show any specific results beyond $d>3$. In addition, published results for this strategy are lacking. In case of GA and ACA, the results are further optimized by a compaction algorithm that attempts to merge the rows of the produced test suite to obtain better optimized results. However, these results do not represent the actual efficiency of the algorithm. As shown in the other tables, CS can generate results beyond $t>3$.

From the results shown in Tables 3 to 8 we observed that IPOG and IPOG-D perform well in all configurations. For the TCAS system in Table 9, IPOG succeed to generate optimized results most of them time. However, IPOG and IPOG-D fail to generate optimized results in most of the tables. Notably, IPOG can achieve better results in case of mixed variables (i.e., when the levels of the input factors are unequal). CTE-XL generates satisfactory results; however it is unable to generate competitive results in some cases. In addition, CTE-XL is unable to generate results beyond $d>3$ as can be noted in Table 3.

Aside from CS, PSO generates better results for most configurations. Similarly, CS could achieve better results for most configurations and could achieve better results compared with PSO. Notably, the size of the CA depends on the values and degrees of combinations, which can be interpreted by the equation of the growth of size that is published in the literature $O(v^t \log k)$ [59].

The results also showed that even when the CS could not generate a better size, it could be the second best performing algorithm. As mentioned previously, deriving a strategy that can generate best results all the time is practically impossible because of the NP-completeness of the problem itself. However, the best strategy can be observed when it can generate small results for most cases.

### 6.2. *Performance evaluation*

In this section, we evaluate the performance of the strategy through the generation time. In addition, we compare the performance of the proposed strategy with other publicly available strategies and tools. The comparison considers only strategies that are publicly available for download and are implemented in the same environment to ensure a fair comparison of generation time. The following strategies are considered for implementation: Jenny, PICT, TConfig, ITCH, TVG, CTE-XL, IPOG-D, and IPOG. Three sets of experiments are conducted. For each experiment, the best size is reported with its generation time for deterministic strategies. In addition, the best and average times of generation are reported for strategies that depend on some degree of randomness. Tables 10 to 12 show the experimental results.

Table 10. Test sizes and execution times for seven input factors, each having three levels, when $2 \leq d \leq 6$

| | | | | | | $v=3$ | | | | | | |
|---|---|---|---|---|---|---|---|---|---|---|---|---|
| | | Jenny | TConfig | ITCH | PICT | TVG | CTE-XL | IPOG-D | IPOG | PSO | CS | |
| | $d$ | N/Time | N/Time | N/Time | N/Time | N/Time | N/Time | N/Time | N/Time | N/Time | N/Time | Avg/Time |
| | 2 | 16/0.37 | 15/029 | 15/17.5 | 16/0.62 | 15/0.22 | 16/0.26 | 18/**0.19** | 17/0.443 | 15/0.21 | **14**/1.68 | 15.2/1.88 |
| $k=7$ | 3 | 51/0.57 | 55/1.86 | **45**/37.85 | 51/0.98 | 55/0.57 | 54/2.55 | 63/0.36 | 57/0.614 | 50/4.21 | 50/**0.133** | 52.4/0.18 |
| | 4 | 169/**0.62** | 166/18.5 | 216/42.62 | 168/1.46 | 167/0.82 | NS | NS | 185/1.357 | **155**/11.32 | 156/3.30 | 157.2/12.50 |
| | 5 | 458/1.91 | 477/198.52 | NS | 452/2.27 | 464/4.602 | NS | 735/**0.86** | 608/2.264 | 441/41.05 | **440**/13.43 | 439.2/15.40 |
| | 6 | 1087/2.58 | 921/1157.8 | NS | 1015/11.524 | 1016/11.524 | NS | 1548/**1.18** | 1281/3.97 | 977/105.59 | **963**/20.41 | 970.2/22.40 |

Table 11. Test sizes and execution times for input configuration when $4 \leq k \leq 10$, each factor having three levels with $d=3$

| | | | | | | $v=3$ | | | | | | |
|---|---|---|---|---|---|---|---|---|---|---|---|---|
| | | Jenny | TConfig | ITCH | PICT | TVG | CTE-XL | IPOG-D | IPOG | PSO | CS | |
| | $k$ | N/Time | N/Time | N/Time | N/Time | N/Time | N/Time | N/Time | N/Time | N/Time | N/Time | avg/Time |
| | 4 | 34/0.08 | 32/0.17 | **27**/32.16 | 34/0.14 | 34/0.17 | 34/0.75 | **27**/**0.04** | 39/0.27 | **27**/0.17 | 29/0.15 | 29.4/0.20 |
| $d=3$ | 5 | 40/**0.12** | 40/0.25 | 45/37.42 | 43/0.45 | 41/0.21 | 43/1.44 | 49/**0.12** | 43/0.34 | **39**/1.739 | **39**/0.13 | 39.4/0.20 |
| | 6 | 51/0.47 | 48/0.67 | **45**/37.62 | 48/0.83 | 49/0.48 | 52/1.96 | 49/**0.12** | 53/0.58 | **45**/2.25 | **45**/0.39 | 46.2/0.55 |
| | 7 | 51/0.57 | 55/1.86 | **45**/37.85 | 51/0.98 | 55/0.57 | 54/2.55 | 63/**0.36** | 57/0.614 | 50/4.21 | 48/0.44 | 49.6/0.58 |
| | 8 | 58/0.73 | 58/2.48 | **45**/38.37 | 59/1.3 | 60/1.251 | 63/2.85 | 63/**0.49** | 63/0.98 | 54/7.15 | 55/2.21 | 55.2/2.45 |
| | 9 | 62/0.82 | 64/3.32 | 75/52.4 | 63/2.76 | 64/1.1812 | 66/4.65 | 71/**0.111** | 65/1.36 | **58**/9.03 | 60/2.39 | 60.8/3.25 |
| | 10 | 65/1.16 | 68/6.71 | 75/52.67 | 65/2.94 | 68/2.414 | 71/5.9 | 71/**0.111** | 68/1.92 | **62**/13.27 | 64/3.15 | 66.2/4.10 |

Table 12. Test sizes and execution times for input configuration when $k=7$, each factor having levels $2 \leq v \leq 6$ with $d=3$

| | | | | | | $k=7$ | | | | | | |
|---|---|---|---|---|---|---|---|---|---|---|---|---|
| | | Jenny | TConfig | ITCH | PICT | TVG | CTE-XL | IPOG-D | IPOG | PSO | CS | |
| | $v$ | N/Time | N/Time | N/Time | N/Time | N/Time | N/Time | N/Time | N/Time | N/Time | N/Time | avg/Time |
| | 2 | 14/0.18 | 16/0.68 | 13/25.6 | 15/0.37 | 15/0.22 | 15/0.32 | 14/**0.15** | 19/0.93 | 13/0.32 | **12**/6.50 | 13.8/7.80 |
| $d=3$ | 3 | 51/0.57 | 55/1.86 | **45**/37.85 | 51/0.98 | 55/0.57 | 54/2.55 | 63/**0.36** | 57/0.614 | 50/4.21 | 50/0.44 | 51.2/0.85 |
| | 4 | 124/1.31 | **112**/4.72 | **112**/93.4 | 124/1.06 | 134/0.95 | 136/5.7 | **112**/**0.43** | 208/0.97 | 116/21.34 | 118/1.49 | 118.8/1.85 |
| | 5 | 236/2.43 | 239/17.53 | **225**/114.5 | 241/1.9 | 260/2.15 | 267/20.5 | 292/**0.95** | 275/2.175 | **225**/35.6 | 233/5.52 | 235.2/6.25 |
| | 6 | **400**/3.85 | 423/84.58 | 1177/585.7 | 413/3.74 | 464/4.458 | 467/55.6 | 532/**1.23** | 455/3.514 | 425/183.56 | 403/12.59 | 410.2/14.50 |

The results in Tables 10 to 12 show the time of test suite generation for a specific configuration along with the size of the test suite. In addition, the best and average sizes for each case are reported because they depend on the randomness of their algorithms up to certain level. Considering that the other strategies are deterministic, only the best results are given. The best generation times are shown in shaded numbers and the best sizes are shown in bold.

The results in Tables 10 and 12 indicate that the generation time increased exponentially with respect to the values and the combination degrees. Notably, the computational strategies generate the test suites faster than the metaheuristic strategies. However, the computational strategies fail to generate better results in terms of size. In this comparison, CS can generate test suites with better performance than PSO. However, IPOG-D can generate test suites with better performance than IPOG and Jenny. This delay in performance is due to the iterative loop applied to achieve optimum fitness values. Despite this situation, IPOG and Jenny can still generate test suites with asymptotic times. Notably, more weight is given to generating a smaller size for the test suites than the generation time. These rules can change depending on the type of the application.

### 6.3. *Effectiveness evaluation through an empirical case study*

An artifact program is selected as the object of the empirical case study. The program is used to evaluate the personal information of new applicants for officer positions. The program consists of various GUI components that represent personal information and criteria to convert them to a weighted number. Each criterion has an effect on the final result, which decides the rank and monthly wage of the officer. The final number is the resulting point. The program is selected because it has a nontrivial code base and different configurations. Figure 10 shows the main window of the program.

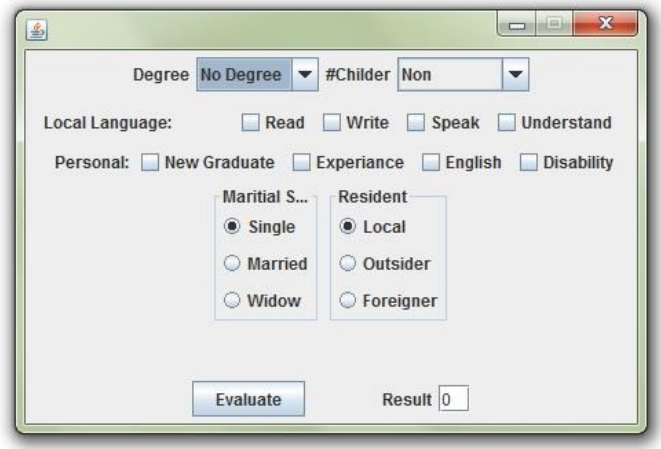

Figure 10. Main window of the empirical study program

The program regards different configurations as input factors. Each input factor has different levels. For example, the user can choose "No Degree," "Primary," "Secondary," "Diploma," "Bachelor," "Master," and "Doctorate" levels for the "Degree" factor. Table 13 summarizes the factors and levels for the program.

Table 13. Summary of the input factors and levels for the case study program

| No. | Factors | Levels |
|---|---|---|
| 1 | Degree | [No Degree, Primary, Secondary, Diploma, Bachelor, Master, Doctor] |
| 2 | children | [non, 1, 2, 3, 4, More_than_4] |
| 3 | read | [checked, unchecked] |
| 4 | write | [checked, unchecked] |
| 5 | speak | [checked, unchecked] |
| 6 | understand | [checked, unchecked] |
| 7 | New graduate | [checked, unchecked] |
| 8 | Experience | [checked, unchecked] |
| 9 | English | [checked, unchecked] |
| 10 | Disability | [checked, unchecked] |
| 11 | Marital Status | [Single, Married, Widow] |
| 12 | Resident | [Local, outsider, Foreigner] |

To this end, the input configuration of the program can be represented by one factor with sevenlevels, one factor with six levels, eight factors with two levels each, and two factors with three levels each. Thus, this input configuration can be notated in an MCA notation as MCA ($N$; $d$, $7^1\ 6^1\ 2^8\ 3^2$). We need 96,768 test cases to test the program with exhaustive configuration testing. In this study, a combinatorial test suite is generated by considering the input configuration to minimize the number of test cases. Table 14 shows the size of each test suite, considering the combination degree.

Table 14. Size of the test suite used for the case study

| Comb. Degree (d) | Test suite size |
|---|---|
| 2 | 42 |
| 3 | 136 |
| 4 | 446 |
| 5 | 1205 |
| 6 | 2886 |

The program was injected with various types of mutations (faults). using MuClipse [76] to verify the effectiveness of the proposed strategy. MuClipse is a mutation injection software that uses muJava as mutation tool. MuClipse creates various types of faults within the original program to test the effectiveness of the generated test suites in detecting these faults.

In general, mutation testing has two advantages on the test suites obtained from the strategy. The first is that it verifies the contribution of different methods and variables defined in the class on the calculation process within the class. The second is that it determines if any similar behavior or reaction exists between the test cases. Deriving similar test cases and reducing the number of cases used for the final test suite are important. The used mutation test was intended to verify and improve the test suite. This is done by detecting and locating similar behavior in the resulted test suite obtained from the combinatorial test suite.

The verification step is carried out by capturing test cases with similar reflexes or behavior (fail or pass) towards different mutation injected in the software. The operation includes two phases. First, is grouping the tests which have the same number of failed and pass cases. Then, matching the reflex patterns of the injected mutations for those similar test cases obtained from the first step. This step is very important, because the combinatorial test suite cannot guarantee that the obtained test cases can have different functional behavior despite having different values.

As shown in Table 14, when the combination degree is 2, 42 test cases were generated from the optimization algorithm, which covers the entire code. muJava generated 278 mutation classes, which are then reduced to 70 mutation classes as a result of similarity in the mutation concept that generates the same effect. Figure 11 shows the reaction of these test cases to the 70 mutation classes.

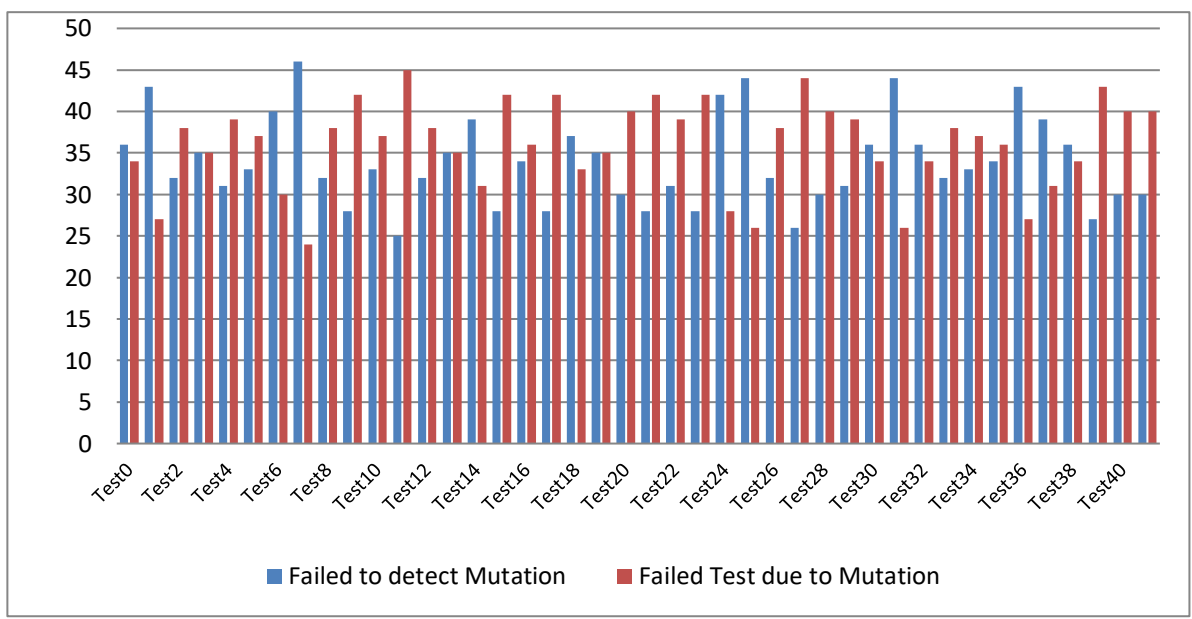

Figure 11. Reaction of the test cases with the configuration for the number of mutations detected when $d=2$.

The blue strips in Figure 11 represent the number of mutation classes that achieved a correct result. In this case, the test case was not affected by the injected mutation because the mutation has no effect on the class calculation and final result. By contrast, the red strips represent the number of failed test cases that resulted from the effect of the injected mutation. In this case, the mutation has a direct effect on the calculated result and thus achieved an incorrect result. In this study, when $d=2$, 12 faults were not detected during the 42 tests.

The number of failed test cases with various mutation classes was used to determine the test cases that have the same response. The cases that have the same number of failed tests were compared to detect any behavior similarity toward the mutations. From the obtained results, test cases 22 and 29 exhibited the same response for all mutation classes. As a result, test case 29 is an excess to the test cases and can be deleted. Meanwhile, the remaining test cases responded differently to the mutation classes and are thus retained.

When the combination degree is 2, 135 test cases are obtained from the program testing strategy. Figure 12 shows the reaction to the same 70 mutation classes used when the combination degree is 2.

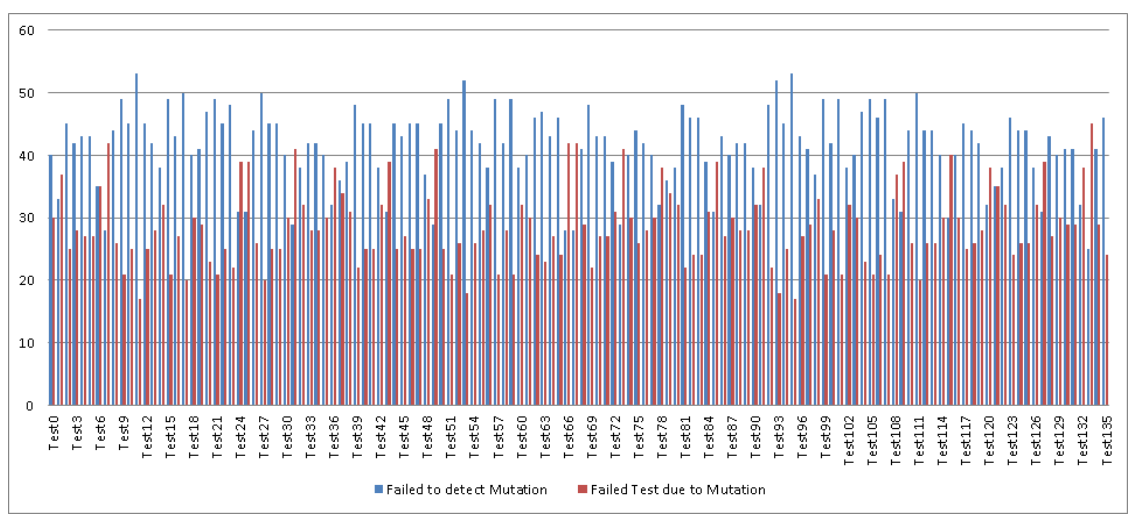

Figure 12. Reaction of the test cases with the configuration for the number of mutations detected when $d=3$

As shown in Figure 12, the 136 test cases were applied to verify similarity in the same manner as in the previous case. The test cases that have the same response to the mutation were deleted to further optimize the final test suite. Notably, many tests could not detect many mutations at once. However, the overall test cases successfully detected all the mutations, including the 12 faults that were not detected by the test suite with a combination degree of 2. The higher combination degree (i.e., the test suites for $d>3$) was also able to detect the faults. However, as far as all the faults that were detected by the test suite of $d=3$, the results were not reported in this study to avoid redundancy.

## 7. Threats to validity

This study encountered different threats to validity, as in other studies. Considerable attention should be focused on reducing these threats by designing and running different experiments. However, we need to address the threats to validity. First, given the lack of results for the metaheuristic algorithms, we need to conduct more experiments for further evaluation to determine the strengths and weaknesses of different algorithms. Second, only one program is used for the case study. Although the program is an ideal artifact for functional testing, more case studies and evidence can show the effectiveness of the strategy. In addition, the faults that were injected in the current program could be detected when the test suite has a combination degree that is equal to 3. However, other types of faults within the same case study can be detected by a higher combination degree.

## 8. Conclusion

In this paper, we propose a strategy for combinatorial test suite generation using CS. The strategy is applicable for functional testing activities in which the internal structure of the code is not considered. CS has recently been proven to be an effective optimization algorithm for NP-complete problems. An extensive evaluation with different benchmarks and experimental cases has been presented to determine the strengths and weaknesses of the proposed strategy. The evaluation results showed that using the CS to optimize the combinatorial test suites could generate better results most of the time compared with its counterpart strategies. A real-world case study is used to evaluate the effectiveness of the test suite generated by the strategy. The strategy proved its effectiveness in detecting faults in programs by using the functional testing approach.